\documentclass[prb,superscriptaddress,preprint,floatfix]{revtex4}
\usepackage{graphicx}
\usepackage{color}
\usepackage{bm,amsmath,amssymb,enumitem,mathtools}
\usepackage{multirow}
\usepackage{ulem} 
\usepackage{float}
\usepackage{soul}
\usepackage{cancel}
\usepackage{inputenc}

\begin{document}

\title{A new look at the temperature-dependent properties of the antiferroelectric model PbZrO$_{3}$: an effective Hamiltonian study}
\author{Kinnary Patel}
\affiliation{Physics Department and Institute for Nanoscience and Engineering, University of Arkansas, Fayetteville, Arkansas 72701, USA}

\author{Bin Xu}
\affiliation{Institute of Theoretical and Applied Physics, Jiangsu Key Laboratory of Thin Films, School of Physical Science and Technology, Soochow University, Suzhou 215006, China}

\author{Sergey Prosandeev}
\affiliation{Physics Department and Institute for Nanoscience and Engineering, University of Arkansas, Fayetteville, Arkansas 72701, USA}

\author{Romain Faye}
\affiliation{Universit\'e Paris-Saclay, CNRS, CentraleSup\'elec, Laboratoire SPMS, 91190 Gif-sur-Yvette, France}

\author{Brahim Dkhil}
\affiliation{Universit\'e Paris-Saclay, CNRS, CentraleSup\'elec, Laboratoire SPMS, 91190 Gif-sur-Yvette, France}

\author{Pierre-Eymeric Janolin}
\affiliation{Universit\'e Paris-Saclay, CNRS, CentraleSup\'elec, Laboratoire SPMS, 91190 Gif-sur-Yvette, France}

\author{Laurent Bellaiche}
\email[Corresponding author:\,]{laurent@uark.edu}
\affiliation{Physics Department and Institute for Nanoscience and Engineering, University of Arkansas, Fayetteville, Arkansas 72701, USA}

\begin{abstract}
A novel atomistic effective Hamiltonian scheme, incorporating an original and simple bilinear energetic coupling, is developed and used to investigate the temperature dependent physical properties of the prototype antiferroelectric PbZrO$_{3}$ (PZO) system. This scheme reproduces very well the known experimental hallmarks of the complex Pbam orthorhombic phase at low temperatures and the cubic paraelectric state of \textit{Pm$\bar{3}$m} symmetry at high temperatures. Unexpectedly, it further predicts a novel intermediate state also of \textit{Pbam} symmetry, but in which anti-phase oxygen octahedral tiltings have vanished with respect to the \textit{Pbam} ground state. Interestingly, such new state exhibits a large dielectric response and thermal expansion that remarkably agree with previous experimental observations and the x-ray experiments we performed. We also conducted direct first-principles calculations at 0K which further support such low energy phase. Within this fresh framework, a re-examination of the properties of PZO is thus called for.
\end{abstract}

 
\maketitle

\section{Introduction}

Antiferroelectrics (AFEs) form an important class of materials that are characterized by antipolar arrangement of dipoles.  Because of the various attractive functionalities provided by these materials, there is a growing interest in their use in technological applications, in particular for energy storage \cite{XHao2013AdvDie,BMa2009-1,BMa2009-2,XHao2013AppPhys,BPeng2015}.  PbZrO$_{3}$ (PZO) is the prototypical antiferroelectric (AFE) perovskite,  and AFE archetype, and its characteristics have been studied since the 1950s. Recent research activities are aimed to better understand its properties \cite{Jorge2014,SergeyPZO2014,JHlinka2014,ECockayne2000,TagantsevPZO,ABussmann2013,BKMani2015}; however,
despite intensive  investigations, PZO  is still puzzling regarding several issues, including the origin and  complex nature of its ground state and possible existence of intermediate phases before reaching its paraelectric high temperature state \cite{HLiu2011,XTan2011}. PZO exhibits a cubic perovskite structure of $Pm\bar{3}m$ symmetry
at high temperatures and an antipolar orthorhombic ground state below a critical temperature, T$_c$, of about 505 K, which has the space group \textit{Pbam} \cite{HFujishita1982,HFujishita1984}. 
This \textit{Pbam} ground state \cite{Jorge2014,SergeyPZO2014,JHlinka2014,ECockayne2000,TagantsevPZO,BKMani2015} consists of three structural distortions in terms of phonon  mode instabilities of the cubic parent phase. The first one is a strong $R^{+}_{4}$ soft phonon mode associated with the zone boundary $\frac{2 \pi}{a} (\frac{1}{2},\frac{1}{2},\frac{1}{2})$  k-point of the cubic first Brillouin zone, where $a$ is the lattice constant of the five-atom cubic perovskite cell.  This mode characterizes an anti-phase tilting of the oxygen octahedra about the [110] direction in the perovskite lattice. A second mode is the $\Sigma_{2}$ mode and is indexed by the $\frac{2 \pi}{a} (\frac{1}{4},\frac{1}{4},0)$  k-point. The $\Sigma_{2}$ mode consists of  complex antipolar displacements of Pb ions along [110] accompanied by  some oxygen displacements resulting in an unusual tilting pattern about the [001] pseudo-cubic axis. There  is also a third mode contributing to the \textit{Pbam} ground state, but that is  rather weak. It has  the $S_{4}$ symmetry and  is associated with the $\frac{2 \pi}{a} (\frac{1}{4},\frac{1}{4},\frac{1}{2})$ k-point \cite{BJaffe1961}.  It has first been assumed that a trilinear coupling between these three modes, {\it i.e.} $R^{+}_{4}$, $\Sigma_{2}$, and $S_{4}$ modes, plays  an important role to stabilize PZO’s \textit{Pbam} ground state (see, {\it e.g.}, Refs.[\onlinecite{SergeyPZO2014,JHlinka2014}]).  However, first-principles studies of Ref.[\onlinecite{Jorge2014}] revealed that the contribution of this trilinear term to the total energy is small compared to the energy gain of the ground state with respect to the cubic phase \cite{Jorge2014}. Recently, a theoretical framework has rather suggested  that another trilinear coupling term, which is similar to flexoelectricity but that involves gradients of the octahedral tilt modes rather than strain, is responsible for the ground state of PZO with \textit{Pbam} symmetry \cite{MStengeltiltgradient}. 
In contrast, a recent experimental work using resonant ultrasound spectroscopy has rather suggested biquadratic and higher order terms couplings through strains between $R$ and $\Sigma$ modes \cite{MACarpenter2022}. Remarkably, Ref.[\onlinecite{Kinnaryuw}] discovered a  novel and simpler atomistic energy which only {\it bi-linearly} couples the A-cation displacements and oxygen octahedral tilting (also known as antiferrodistortive distortion (AFD)) in ABO$_{3}$ perovskites and which straightforwardly provided an unified description of many antiferroelectric and incommensurate perovskites, as well as ferrielectrics \cite{BinTanPZO}. This finding  therefore raises the question if  such previously overlooked and simple bi-linear coupling  is, in fact, the main contributor to stabilize the \textit{Pbam} ground state of  PZO.

Moreover, new low-energy phases were recently predicted by first-principles calculations in PZO, such as those of $Ima2$ and $Pnam$ symmetries containing  30 atoms and 80 atoms per primitive cell, respectively \cite{SergeyPZO2014, Hugo2021, BinBaker2021}. The energy difference between these phases is very small and dramatically depends on the pseudopotentials and other computational details, further emphasizing that PZO is rather challenging to understand and to correctly simulate. PZO has another unsettled related question regarding a possible intermediate phase for temperatures in-between the cubic paraelectric $Pm\bar{3}m$ state and the antiferroelectric  orthorhombic \textit{Pbam} ground state. Experimentally, such an intermediate phase is sometimes stabilized for a narrow temperature range between the AFE and the cubic phase but still, the structure of the intermediate phase is not really fully known \cite{Sawaguchi1951PZOTc,Tennery1965,Tennery1966,Goulpea1967,Tanaka1982,Fujishita1992,Liu2018}. Depending on the crystal growth conditions, intrinsic or intentional chemical doping/defects may extend the temperature range in which the intermediate phase develops and/or causes the appearance of a second intermediate phase showing isothermal, i.e., time-dependent, transition process \cite{HLiu2011,DKajewski2022}. A recent first-principles study by Xu. {\it et al.} suggested  possible candidates for intermediate phases of PZO \cite{orderdisorderBin2019}, including one consisting of a dynamical average between the rhombohedral ferroelectric $R3c$ phase and antiferroelectric \textit{Pbam} phases. Involvement of ferroelectric rhombohedral phases and/or intermediate state was also indicated  (1) in a high temperature X-ray diffraction study proposing the existence of a ferroelectric rhombohedral phase on heating the PZO crystal \cite{Tennery1966}; and (2) by Tangantsev {\it et. al.} \cite{TagantsevPZO}, who  studied the lattice dynamics of PZO from X-ray and Brillouin scattering and showed that PZO exhibits an intermediate phase on heating the crystal. One can also find in the literature other possibilities for that intermediate phase, such as one associated with lattice distortion corresponding to the M-point (k = $\frac{2 \pi}{a}(\frac{1}{2}, \frac{1}{2},0)$) of the first Brillouin zone -- as similar to the phase obtained in Ref [{\onlinecite{HFujishita1984}]. Note also that based on several experimental techniques, some polar clusters have been proposed to co-exist in the paralectric phase, suggesting that the purely paraelectric state is only achieved above ~593 K, i.e., far above T$_C$ (see Ref. \onlinecite{JHKo2013Modesoftintermediatephase2013}).

The goal of this article is two-fold. First of all, to demonstrate, via the development of a novel {\it ab-initio} effective Hamiltonian (H$_\textrm{eff}$), that the bi-linear coupling of Ref.  \cite{Kinnaryuw} can indeed lead to the stabilization of the complex antiferroelectric  orthorhombic \textit{Pbam} ground state in PZO; Secondly, to use such novel atomistic scheme to predict that there is indeed an intermediate state in PZO being in-between the known Pm${\bar 3}$m and \textit{Pbam} states, but which is a state that has never been previously mentioned  in the literature -- to the best of our knowledge  and that shows instabilities allowing to better understand the reported experimental observations. Effective Hamiltonian calculations  yield that such intermediate state can be thought as originating from the known \textit{Pbam} ground state but when removing the $R^{+}_{4}$ phonon mode (and also the $S_{4}$  mode). Such state, which is further confirmed here to be of low energy by conducting additional direct ab-initio calculations, is coined here 
the $\Sigma$ phase, due to the predominance of the $\Sigma_{2}$ mode. It also has the  \textit{Pbam} symmetry, therefore resulting in an isostructural transition in PZO with temperature when going from that phase to the ground state under cooling. Interestingly, the H$_\textrm{eff}$ computations also provide  an intermediate state that possesses (i) a large dielectric response, as it is experimentally known in PZO for temperatures above the transition towards the ground state\cite{RomanFayeRef11,RomanFayeRef14,RomanFayeRef46,RomanFayeRef48,RomanFayeRef55,GShirane1951DEResponse,RomanFayeRef69,RomanFayeRef70,RomanFayeRef80,RomanFayeRef81}); and (ii)  a thermal expansion that agrees well with our measurements.These calculations also yield pseudocubic lattice parameters that are all close to each other in this intermediate state, which indicates that this novel \textit{Pbam} $\Sigma$ state can be thought to be  a cubic phase in disguise and thus may explain why it may have been missed up to now.  This $\Sigma$ state is also found to be very close in free energy to a tetragonal ferroelectric P4mm phase or other ferroelectric states that we believe could be easily triggered through constrains such as chemical doping and defects, as well as the application of external electric or mechanical fields, and, therefore explain many of the experimental observations.
 
 This article is organized as follows. Section II reports details about theoretical and experimental methods developed and/or used here. Section III provides and discusses the results, while Section IV summarizes the article.

\section{Methods}

\subsection{Effective Hamiltonian} 

As indicated above, an effective Hamiltonian, H$_{\text{eff}}$, is developed in the aim of understanding and modeling finite-temperature properties of PbZrO$_3$ bulk. This H$_{\text{eff}}$ has the following  degrees of freedoms: (1) the local soft modes $\bm{u}_i$ in each 5-atom cells $i$,  which are proportional to the electric dipoles of that cell \cite{Zhong} and that are centered on Pb ions here; (2) the $\bm{\omega}_i$ pseudo-vectors that describes oxygen octahedral tiltings  in the unit cells $i$ \cite{Igor2006} and that are centered on Zr ions; $\bm{\omega}_i$ is such as its direction is the axis about which the oxygen octahedron of cell $i$ rotates and its magnitude is the angle (in radians) of such rotation; (3)  the $\{ \eta_H\}$ homogeneous strain tensor for which the zero value of its diagonal elements (in Voigt notation) $ \eta_{H,1}= \eta_{H,2}=\eta_{H,3}$ is associated with the calculated first-principles-derived lattice constant of the paraelectric cubic state of PbZrO$_3$ at 0 K ;  and (4) $\bm{v}_{i}$ vectors that quantify the inhomogeneous strain at each 5-atom cell $i$  ~\cite{Zhong}, and that are centered  on Zr ions here. 

Following Refs.[\onlinecite{NaNbO3Heff,CsPbI3Heff,BaCeO3Heff}], the total internal energy contains two different main energies:
\begin{align}\label{eq:etot}
E^{\text{tot}} &= E^{\text{FE}}(\{\bm{u}_i\},\{\eta_l\})+E^{\text{tilt}}(\{\bm{\omega}_i\},\{\bm{u}_i\},\{\eta_l\}) 
\end{align} 
where $E^{\text{FE}}$  describes energetics associated with local modes, elastic variables and their interactions, while $E^{\text{tilt}}$ characterizes energetics involving oxygen octahedral tilts and their couplings with local modes and the total $\{\eta_l\}$ strain (that is the sum of inhomogeneous and inhomogeneous strains).  

As indicated in Ref.[\onlinecite{Zhong}], $E^{\text{FE}}$  can be decomposed into five terms: 
\begin{align}
\nonumber E^{\text{FE}}  & = E^{\text{self}}(\{\bm{u}_i\})+E^{\text{dpl}}(\{\bm{u}_i\})+E^{\text{short}}(\{\bm{u}_i\}) \\  
&+E^{\text{elas}}(\{\eta_l\})+E^{\text{int}}(\{\bm{u}_i\},\{\eta_l\}) ,
\end{align}
where $E^{\text{self}}$ is the local mode self energy, $E^{\text{dpl}}$ pertains to the long-range dipole-dipole interaction, $E^{\text{short}}$ represents the short-range interactions between local modes that go beyond dipole-dipole interactions, $E^{\text{elas}}$ denotes the elastic energy, and $E^{\text{int}}$ describes the interaction between elastic variables and local modes. As proposed in Ref.[\onlinecite{Zhong}], these five energies can be written as follows:
\begin{align} 
\nonumber E^{\text{self}} &= \sum_{i} \{ \kappa_2 u^2_i+\alpha u^4_i+\gamma (u^2_{ix}u^2_{iy}+u^2_{iy}u^2_{iz}+u^2_{ix}u^2_{iz}) \} \\ \nonumber
E^{\text{dpl}} &= \frac{Z^{*2}}{\epsilon_\infty}\sum_{i<j}\frac{\bm{u}_i\cdot\bm{u}_j-3(\hat{\bm{R}}_{ij} \cdot\bm{u}_i)(\hat{\bm{R}}_{ij} \cdot\bm{u}_j)}{R^3_{ij}} \\ \nonumber
E^{\text{short}} &= \sum_{i\neq j}\sum_{\alpha \beta} J_{ij\alpha \beta}u_{i\alpha}u_{j\beta} \\ \nonumber
E^{\text{elas}} &= \frac{N}{2}B_{11}(\eta^2_1+\eta^2_2+\eta^2_3)+NB_{12}(\eta_1\eta_2+\eta_2\eta_3+\eta_3\eta_1) \\ \nonumber
&+\frac{N}{2}B_{44}(\eta^2_4+\eta^2_5+\eta^2_6) \\ 
E^{\text{int}} &= \frac{1}{2}\sum_i\sum_{l\alpha\beta}B_{l\alpha\beta}\eta_l(\bm{R}_i)u_{\alpha}(\bm{R}_i)u_{\beta}(\bm{R}_i),
\end{align}
with $\bm{R}_{ij}=\bm{R}_i-\bm{R}_j$ for which $\bm{R}_i$ and  $\bm{R}_j$ are the lattice vectors locating sites $i$ and $j$, respectively. $\alpha $ and $\beta $ denote Cartesian components along the [100], [010] and [001] directions, that are chosen to span the x-, y-, and z-axes, respectively.  The  $i$ index runs over all the Pb ions. For any given  Pb-site $i$, the $j$ index in $E^{\text{dpl}}$ runs over all the other Pb ions.  In contrast, the $j$ index in  $E^{\text{short}}$ ``only'' runs over the first, second and third nearest Pb neighbors of the Pb ion located at the site  $i$.

Furthermore, the  $J_{ij\alpha \beta}$ coefficients in $E^{\text{short}}$ can be written in the following way for the different nearest neighbor (NN) interactions:
\begin{align}
\nonumber \text{first NN:} \; J_{ij\alpha \beta}=&(j_1+(j_2-j_1)|\hat{\bm{R}}_{ij,\alpha}|)\delta_{\alpha\beta};\\
\nonumber \text{second NN:} \; J_{ij\alpha \beta}=&(j_4+\sqrt{2}(j_3-j_4)|\hat{\bm{R}}_{ij,\alpha}|)\delta_{\alpha\beta}\\
\nonumber         &+2j_5\hat{\bm{R}}_{ij,\alpha}\cdot\hat{\bm{R}}_{ij,\beta}(1-\delta_{\alpha\beta});\\
 \text{third NN:} \; J_{ij\alpha \beta}=&j_6\delta_{\alpha\beta}+3j_7\hat{\bm{R}}_{ij,\alpha}\cdot\hat{\bm{R}}_{ij,\beta}(1-\delta_{\alpha\beta}),
\end{align}
where $\delta$ is the Kronecker symbol and $\hat{\bm{R}}_{ij,\alpha}$ represents the $\alpha$-component of $\bm{R}_{ij}/R_{ij}$. A detailed explanation of the  $j_{3}$ coefficient is mentioned in Ref [{\onlinecite{Zhong}]}. It represents the strength (and sign)  of a specific intersite interaction  between second nearest neighbor for the local mode.

Regarding the second main energy of Eq. (1), we generalize that of  Refs.[\onlinecite{NaNbO3Heff,CsPbI3Heff,BaCeO3Heff,SergeyNano}]  by writing:

\begin{align}\label{eq:afd}
\nonumber E^{\text{tilt}} (\{ \omega _{i} \}, \left\{u_{i} \right\},\{\eta_l\}) = &\sum_i [\kappa_A\omega^2_i+\alpha_A\omega^4_i+\gamma_A(\omega^2_{ix}\omega^2_{iy}+\omega^2_{iy}\omega^2_{iz}+\omega^2_{ix}\omega^2_{iz})] \\
\nonumber +&\sum_{ij}\sum_{\alpha\beta}K_{ij\alpha\beta}\omega_{i\alpha}\omega_{j\beta} +\mathop{\sum }\limits_{i} \mathop{\sum }\limits_{\alpha } K'  \omega _{i,\alpha }^{3} \; (\omega _{i+\alpha ,\alpha } +\omega _{i-\alpha ,\alpha } ) \\
\nonumber +&  \sum_i\sum_{\alpha\beta}C_{l\alpha\beta}\eta_l(i)\omega_{i\alpha}\omega_{i\beta} \\
\nonumber +&\mathop{\sum }\limits_{i,j} \mathop{\sum }\limits_{\alpha ,\beta } {\rm D}_{ij,\alpha \beta } u_{j,\alpha } \omega _{i,\alpha } \omega _{i,\beta }  +  \sum_{i,j}\sum_{\alpha\beta\gamma\delta}E_{\alpha\beta\gamma\delta}\omega_{i\alpha}\omega_{j\beta}u_{j\gamma}u_{i\delta} ~\\
 +&\mathop{\sum }\limits_{i,j} \mathop{\sum }\limits_{\alpha ,\beta } {\rm F}_{ij,\alpha \beta } \omega _{i,\alpha } u_{j,\beta }  
\end{align}

   where the sums over $i$ run over all the Zr sites. The first sum of $E^{\text{tilt}}$ describes the onsite contributions  associated with the oxygen octahedral tilts, as  proposed in Refs.[\onlinecite{Albrecht,Igor2007,Lisenkov,SergeyNano}].  
The second and third terms represent short-range interactions between oxygen octahedral tiltings \cite{SergeyNano,Igor2007} being first-nearest neighbors in the Zr sublattice. Note that $j$ runs over the six Zr sites that are first-nearest neighbors of the Zr site $i$ in the second term while $\omega _{i+\alpha ,\alpha }$ in the third term denotes the $\alpha $-component of the $\omega$ pseudo vector at the site shifted from the Zr site $i$ to its nearest Zr neighbor along the $\alpha$ axis. Note also that the $K_{ij\alpha \beta}$ parameters of this second energy are written as:
\begin{align}
\text{first NN:} \; K_{ij\alpha \beta}=&(k_1+(k_2-k_1)|\hat{\bm{R}}_{ij,\alpha}|)\delta_{\alpha\beta}
\end{align}
Here, $k_{2}$ characterizes the strength and sign of a specific  interaction between first nearest neighbors for the tilting modes (it is the equivalent of $j_2$ shown in Fig. 1 of Ref. [{\onlinecite{Zhong}}] but for the tilting degrees of freedom rather than the local modes). Moreover and following Ref.[\onlinecite{Igor2007}], the fourth term of $E^{\text{tilt}}$ characterizes the interaction between strain and tiltings. As first proposed in Ref.[\onlinecite{SergeyNano}], the fifth term (that depends on the ${\rm D}_{ij,\alpha \beta }$ coefficients) represents a trilinear coupling between oxygen octahedral tiltings and local modes. The sixth energy, which involves the $E_{\alpha\beta\gamma\delta}$ parameters, characterizes bi-quadratic couplings between such tiltings and modes \cite{Igor2007}. Note that the $j$ index runs over the eight Pb ions that are first nearest neighbors of the Zr-site $i$  in these  fifth and sixth terms.  Finally, the seventh term is the novelty here with respect to the energies provided in Refs.[\onlinecite{Albrecht,Igor2007,Lisenkov,SergeyNano}]. It represents a recently discovered energy that couples oxygen octahedral tiltings and local modes in a bi-linear fashion \cite{Kinnaryuw,Reviewcouplingenergies}. It is allowed by symmetry and was proposed to explain the occurrence of complex antiferroelectric, ferrielectric and even incommensurable phases in some materials \cite{Kinnaryuw,Reviewcouplingenergies,BinTanPZO}. Note that the sum over j is about the eight Zr-sites that are nearest neighbors of the Pb-sites i.  It is given, in a less compact form than in Eq. (5), by

\begin{flalign}\label{uwenergy}
\nonumber \Delta E  = 
\nonumber& \mathrm{F_{ii}} \sum_i [(u_{i,x} + u_{i,y} )(-~ \omega_{i100,z}+~ \omega_{i010,z}-~ \omega_{i101,z} +~ \omega_{i011,z} )\\
 +& (-u_{i,x} + u_{i,y} ) (-~ \omega_{i,z}+~ \omega_{i110,z}-~ \omega_{i001,z} +~ \omega_{i111,z} )  + \mathrm{cyclic ~permutation}]
\end{flalign}

where $\mathrm{F_{ii}}$ is a material-dependent constant that characterizes the strength of this coupling. The sum over $i$ runs over all the
5-atom cells of the perovskite structure, and the $x$, $y$, and $z$ subscripts denote the Cartesian components of the ${\bf u}_{i}$
vectors and $\boldsymbol{\omega}_{i}$ pseudo-vectors.  $\omega_{ilmn}$, with l, m or n being 0 or 1, characterizes the $\omega$ pseudo-vectors located at   $a (l \hat{x} + m \hat{y} + n \hat{z}$ from that of site $i$), with $a$ being the 5-atom lattice parameter and $\hat{x}$, $\hat{y}$ and $\hat{z}$ being the unit vectors along the x-, y- and z-axes, respectively.  Note that the Zr site $i$ is located at $-a (\frac{1}{2} \hat{x} + \frac{1}{2} \hat{y} + \frac{1}{2}  \hat{z})$ with respect to the Pb site $i$. Figure 1 schematizes the coupling terms inherent to the first line of Eq. \eqref{uwenergy}. This new coupling term is important in some systems having Pb atoms (Ref [\onlinecite{Kinnaryuw}]), hence relaxors.

All the parameters of  this effective Hamiltonian are provided in Table 1. They are determined by conducting  first-principles calculations  using the local density approximation (LDA) \cite{LDA} within density functional theory.  Cells containing  up to 40 atoms are employed, along with the CUSP code \cite{Laurent3} and the ultrasoft-pseudopotential scheme \cite{David} with a 25 Ry plane-wave cutoff.
Practically, we use  the same pseudopotentials than those of Ref.[\onlinecite{DomenicKS}], and thus consider the following valence electrons: Pb 5d, Pb 6s, Pb 6p, Zr 4s, Zr 4p, Zr 4d, Zr 5s, O 2s and O 2p electrons. It is important to know that these effective Hamiltonian parameters are determined by considering small perturbations with respect to the cubic state, and thus do not involve the full ionic and cell relaxation of any phase \cite{Zhong}. Consequently, the fact that we used LDA to obtain these parameters does not automatically imply that our energetic results for {\it relaxed} structures predicted by the effective Hamiltonian will be closer to those resulting from the direct use of LDA than to those obtained from other functionals such as the  generalized gradient approximation (GGA) in the form of revised Perdew-Burke-Ernzerhof \cite{pbesolPerdew2008} (PBEsol). This 
is  especially true if employing these two functionals within first-principles calculations provides similar results in term of structure but  rather small differences between the total energy of different relaxed phases -- as it is, in fact, known in PbZrO$_3$ bulk \cite{HAramberri2021}.

 This H$_{\text{eff}}$, along with  its parameters, is then used  in Monte-Carlo (MC) computations on $12 \times 12 \times 12$ supercells (which contains 8,640 atoms) for different temperatures. Typically,  40,000 MC sweeps  are conducted for each considered temperature, with the first 20,000 sweeps allowing the system to be in thermal equilibrium and the next 20,000 sweeps being employed to obtain statistical averages. However, near phase transitions, a larger number of MC sweeps is needed to have converged results, namely up to  1 million (with the first half used for reaching thermal equilibrium and the next half to extract statistical averages).
 It is also important to know that a temperature-dependent pressure is added in these Monte-Carlo simulations, with this pressure, P, acting on the strains to produce the following energy: 
 \begin{flalign}
 E_{pres}= P  a^{3} (1+\eta_{H,1}) (1+\eta_{H,2}) (1+\eta_{H,3})
 \end{flalign}

 where  $\eta_{H,i}$, with $i$ = 1, 2 and 3, being the diagonal elements of the homogeneous strain tensor in  Voigt notations \cite{Voigt1910} (note that the shear strains are neglected in Equation (8) because they are either null for the cubic paralectric phase or found to be rather small for the other phases encountered in the simulations).  Such dependence is assumed here to be linear, as proposed in Ref.[\onlinecite{tintePvsT}], and is given, in GPa, by: $P=-5.489655172-0.001034483 \times T$, where $T$ is the temperature in Kelvin  with the two coefficients having been numerically found by trying to reproduce the experimental lattice constants at 10\,K and 300\,K.

In order to determine which structural phases are predicted from the effective Hamiltonian simulations,  the following quantities are extracted from the outputs of the MC simulations at any investigated temperature: 

\begin{eqnarray}
u_{{\bf k},\alpha}=\frac{1}{N} \sum_{i} u_{i,\alpha}  \text{exp}(i {\bf k} \cdot {\bf R}_i) \\
\omega_{{\bf k},\alpha}=\frac{1}{N} \sum_{i} \omega_{i,\alpha}  \text{exp}(i {\bf k} \cdot {\bf R}_i)
\end{eqnarray}
where ${\bf k}$ are vectors belonging to the cubic first Brillouin zone, $\alpha$ denotes Cartesian components and the sums run over all the 5-atom sites. 
Typically, for the local modes, we look at the ${\bf k}$-vectors at the zone-center (for the polarization) but also at the $\Sigma$-point (for complex antipolar displacements associated with the \textit{Pbam} state) that is defined as $\Sigma=(\frac{1}{4},\frac{1}{4},0)$  in $2 \pi/a$ units. Such latter point is also investigated for the tilting of the oxygen octahedra, in addition to the R-point that is given by
 $R=(\frac{1}{2},\frac{1}{2},\frac{1}{2})$ still in $2 \pi/a$ units. Note that a non-zero $\omega_{R,\alpha}$ characterizes antiphase tilting about the $\alpha$-axis while a finite $\omega_{\Sigma,\beta}$ should also happen for the \textit{Pbam} ground state with $\beta$ being different from $\alpha$.

\subsection{Direct First-principles calculations}

We also used direct first-principles calculations to check some unexpected predictions arising from the use of the effective Hamiltonian and that will be discussed in the Results' section.  Practically, we use the generalized gradient approximation (GGA) within the revised Perdew-Burke-Ernzerhof functional (PBEsol) \cite{pbesolPerdew2008}, as implemented in the VASP package \cite{Kresse1999}, for these first-principles computations. The projector augmented wave (PAW) \cite{Blochl1994,Kresse1999} is also applied to describe the core electrons, and we consider the Pb ($5d^{10}$$6s^26p^2$),  Zr ($4s^2$$4p^65s^24d^2$)  and  O  ($2s^2$$2p^4$) as valence electrons  with a 520 eV plane-wave cutoff.

\subsection{Experiments}

The lattice parameters have been determined by X-ray diffraction using  a high resolution diffractometer with a Copper radiation ($\lambda$=Cu$_{\textrm{K}_{\alpha,1}}$=1.5405\,\AA) issued from }an 18\,kW-rotating anode  from room temperature to 750K using a furnace with a resolution better than 0.1\,K. 
The X-ray diagrams were analyzed fitted with a pseudo-tetragonal unit cell due to the overlap of the peaks related to the $a$ and $b$ lattice parameters of the orthorhombic unit cell.
The  lattice parameters at 10\,K and 300\,K have been obtained by Rietveld analysis (Jana software) \cite{Rietveld} on results obtained from neutron diffraction performed at the Laboratoire L\'eon Brillouin (beam line 3T2 , $\lambda$ = 1.2252\,\AA). 
The  lattice parameters obtained at room temperature from neutron diffraction and from X-ray diffraction are in very good agreement. The orthorhombic unit cell parameters (a$_\textrm{o}$, b$_\textrm{o}$, c$_\textrm{o}$) are presented in Fig.(3b) in a pseudo-tetragonal (a$_\textrm{pt}$, c$_\textrm{pt}$)  setting where a$_\textrm{o}=\sqrt{2}$a$_\textrm{pt}$, b$_\textrm{o}\approx 2\sqrt{2}$a$_\textrm{pt}$ and c$_o=2$c$_\textrm{pt}$. At room temperature the orthorhombic cell parameters are a$_\textrm{o}$=5.878\AA, b$_\textrm{o}$=11.783\AA, c$_\textrm{o}$=8.228\AA\, while, at 10\,K, they are a$_\textrm{o}$=5.878\AA, b$_\textrm{o}$=11.784\AA, and c$_\textrm{o}$=8.197\AA.

\section{Results}

\subsection{Results from the H$_{\text{eff}}$ simulations and X-ray diffractions}

Let us now report the predictions of the effective Hamiltonian with the parameters indicated in Table 1 and the total energy described in Equation (1), with the additional feature that 
the second line of Equation (7) is dropped out. This dropping is made because we numerically found that incorporating all the terms of Eq.(7) provides unphysical solutions for tilting arrangements (such as tiltings associated with the $X$-point of the first Brillouin zone, which is given by $(0,0,\frac{1}{2})$ in  $2 \pi/a$ units) while keeping the first eight terms of Eq.(7) gives rise to a selected \textit{Pbam} ground state (Note that dropping all terms of Eq.(7) will not yield such latter ground state). 

 Figure 2(a) reports the temperature evolution of the non-zero tilting-related quantities, namely $\omega_{R,x}=\omega_{R,y}$ and $\omega_{\Sigma,z}$, while Figure 2(b) shows that the only significant local mode quantity  is $u_{\Sigma,x}=u_{\Sigma,y}$, as obtained by heating up the system from the ground state (note that identical results are obtained by cooling down the system, and that the H$_{\text{eff}}$ calculations also provide a $S_{4}$ mode, but for which the weight is rather small, namely less than 0.32\%).
The fact that all these quantities are finite at small temperature does characterize a \textit{Pbam} ground state.

As the temperature increases up to $\simeq$ 650\,K, $u_{\Sigma,x}=u_{\Sigma,y}$ decreases while $\omega_{\Sigma,z}$ is barely affected and  $\omega_{R,x}=\omega_{R,y}$ (anti-phase oxygen octahedra tilts) gets reduced significantly until vanishing through a jump.  A first-order transition to what we call here the $\Sigma$-phase, and that is only characterized by non-zero $u_{\Sigma,x}=u_{\Sigma,y}$ and $\omega_{\Sigma,z}$ (complex tilts along z direction), occurs. 
Such latest phase still has the \textit{Pbam}  symmetry with space group no. 55 (as indicated by the FINDSYM software \cite{FINDSYM,HTStokes2005}), therefore indicating that our predicted transition around 650\,K is isostructural in nature \cite{VRajan2005isostrucutral,Tian2018, Kennedy1999Isostructuralphasetransition}. Note that  isostructural transitions are usually of first-order, which is consistent with the jump  of   $\omega_{R,x}=\omega_{R,y}$  seen here at 650\,K \cite{SergeyPZO2014}.
Interestingly, it is experimentally known that the \textit{Pbam} ground state phase disappears around  511\,K, which is not so far away from our predicted  650\,K -- especially taking into account that effective Hamiltonians have the tendency to not accurately predict transition temperatures (while qualitatively reproducing phase transition sequences, as demonstrated, {\it e.g.}, by Ref.[\onlinecite{Zhong}]). However, we are not aware of any previous work mentioning that  the \textit{Pbam} ground-state phase transforms into another \textit{Pbam} state (that is, the $\Sigma$ phase here)  under heating, therefore making our predictions provocative and novel.
 Within this $\Sigma$ phase, Figs 1a and 1b indicate that both $\omega_{\Sigma,z}$ and $u_{\Sigma,x}=u_{\Sigma,y}$  decrease with temperature and then vanish through a jump at about 1300\,K. A first-order transition to cubic paraelectric $Pm\bar{3}m$ is thus predicted to happen, according to our effective Hamiltonian simulations. 

In order to determine if these rather surprising predictions of the H$_{\text{eff}}$ can be realistic, Fig. 3  reports the computed $\chi_{11}$, $\chi_{22}$ and $\chi_{33}$ diagonal elements of the dielectric  tensor as well as their average $\chi_{av}=\frac{1}{3} (\chi_{11}+\chi_{22}+\chi_{33})$ as a function of temperature.
One can see that $\chi_{33}$ and thus $\chi_{av}$ adopt very large values at the temperatures around which the known  \textit{Pbam} ground state transforms into the novel $\Sigma$ phase, which is consistent with the experimental observation of a large dielectric response around the temperatures at which the known \textit{Pbam} ground state disappears \cite{RomanFayeRef11,RomanFayeRef14,RomanFayeRef46,RomanFayeRef48,RomanFayeRef55,GShirane1951DEResponse,RomanFayeRef69,RomanFayeRef70,RomanFayeRef80,RomanFayeRef81}. Strikingly too, the coefficient $C$ in the fitting of $\chi_{33}$ and $\chi_{av}$ by $\frac{C}{T-T_0}$ is predicted to be 1.81$\times$10$^5$\,K and 1.49$\times$10$^5$\,K, respectively,  in the $\Sigma$ phase, which is of the same order of magnitude than the experimental values ranging between 1.36$\times$10$^5$\,K and 2.07 $\times$10$^5$\,K  (see Refs.[\onlinecite{RomanFayeRef46,RomanFayethesis}] and references therein) for temperatures at which the known \textit{Pbam} ground state has vanished.

Note that we found that the increase of $\chi_{av}$ and $\chi_{33}$ when the temperature gets reduced in the $\Sigma$ phase towards the known \textit{Pbam} state can be thought to originate from the fact that there is a $P4mm$ ferroelectric state (with a polarization along the $z$-axis and no oxygen octahedral tilting)  that is very close in free energy for these temperatures. In fact, reducing the $F_{ii}$ parameter in Table I to a certain critical value of 0.019  makes the intermediate phase in-between the known \textit{Pbam} state and the $Pm\bar{3}m$ cubic state becoming $P4mm$ rather than $\Sigma$. Note also that the large dielectric response shown in Fig. 3 implies that the zone-center mode should soften when approaching the known \textit{Pbam} phase from above, which has been indeed observed \cite{OstapchukPolarandCentralmodePZO2001,TagantsevPZO}.  It is also worth mentioning that the proximity in energy of the ferroelectric P4mm state (or other ferroelectric states) could also explain the experimental electric-field- or defect-driven single ferroelectric P-E loop observed above the known Pbam phase \cite{HLiu2011}, or stabilized under epitaxial strain \cite{LPintilie2007}.

Let us now take a look at the behavior of the pseudocubic lattice parameters as a function of temperature. Figure. 4a reports our predictions from the use of the presently developed effective Hamiltonian (obtained from the strain outputs) while Fig. 4b displays our own experimental results along with those of Ref.[\onlinecite{Fujishita2003}].  One can see that, within the known \textit{Pbam} state, several features of the measurements are well reproduced, namely the $a$ and $b$ pseudocubic lattice constants are basically independent of the temperature while the $c$ pseudocubic lattice parameter significantly  increases when heating PZO. The jump in the $c$ lattice constant  seen in Fig. 4a at around 650\,K confirms the  first-order character of the isostructural transition between the known \textit{Pbam} state and the $\Sigma$ phase.   What is remarkable is that, within the error bars (the range of error bars for the $a$ and $b$ pseudocubic lattice parameter  is from 0.0013 to 0.007 {\AA} for the  whole temperature range, while it is less than 0.0038 {\AA} for the $c$ lattice parameter), the $a$, $b$ and $c$ parameters are identical within our predicted $\Sigma$ phase, which may thus give the (wrong) impression that PZO is cubic if solely focusing on pseudocubic lattice parameters or strains. One can thus think that this $\Sigma$ phase is a cubic state in disguise, when focusing on pseudocubic lattice parameters. In fact, the  predicted temperature evolution of $\frac{1}{3} (a+b+c)$ adopts a thermal expansion  that is very close in the $\Sigma$ and paraelectric  $Pm\bar{3}m$ phases, and that is characterized by an average coefficient of 7.97 $\times$ 10$^{-6}$\,K$^{-1}$, \textendash which is in good agreement with the present measurements giving 7.4$\times$ 10$^{-6}$\,K$^{-1}$ (note that such coefficient is numerically found to be 7.77 $\times$ 10$^{-6}$\,K$^{-1}$
in our $\Sigma$ phase {\it versus} 8.36 $\times$ 10$^{-6}$\,K$^{-1}$ in the simulated  $Pm\bar{3}m$ state). This facts suggests, once again, that the present simulations can be realistic. 
 On the other hand, comparing Fig. 4a and 4b tells us that the jump in the $c$ pseudocubic lattice parameter when the known \textit{Pbam} state disappears is more pronounced in experiments than in our present calculations. Such quantitative discrepancy can be due to the facts that this disappearance happens at higher temperature in the simulations and/or that the 
computations use relatively small $12 \times 12 \times 12$  supercells (implying that the first-order character of that transition can be reduced by the increase in critical temperature and/or finite-size effects).

Interestingly, near the temperatures at which the known $Pbam$ state disappears, it has been observed in Ref [71] (see Fig. 2.9 there) that the $(\frac{1}{2},\frac{3}{2},\frac{1}{2})$  diffractions peak typically associated with antiphase oxygen octahedral  tiltings disappears at a temperature lower than the $(\frac{1}{4},\frac{3}{4},0)$ diffraction peak that typically characterizes antiparallel displacements of Pb associated with the $\Sigma$ k-point. Such features appear to be qualitatively consistent with our predictions shown in Figs 1a and 1b. Note that our predicted $\Sigma$-phase can be of short-range or broken into small domains in grown samples, which may explain why it has not been directly reported and reported yet. It is also interesting to realize that, in addition to a zone-center soft-mode, a central mode, with a frequency being very low, has been experimentally reported for temperatures above the known \textit{Pbam} state \cite{OstapchukPolarandCentralmodePZO2001,TagantsevPZO,JHKo2013Modesoftintermediatephase2013}. Such feature may be representative of dynamical jumps between different $\Sigma$-phases, that are with positive or negative $\omega_{\Sigma,z}$, $\omega_{\Sigma,y}$ and $\omega_{\Sigma,x}$. Such dynamical jumps will render cubic the overall structure of PZO, while not affecting $\frac{1}{3} (a+b+c)$.

\subsection{Results from first-principles calculations}

Let us now check if such $\Sigma$ phase is also seen as a low-energy state within direct first-principles calculations. 
Table 2 reports the energy of such $\Sigma$ phase, but also those of the known \textit{Pbam} and $R3c$ states, choosing the zero of energy to correspond to the cubic paraelectric $Pm\bar{3}m$  state.
One can first see that the known \textit{Pbam} is the state with the lowest energy but by only a small amount of about 1 meV/f.u. with respect to $R3c$. Very interestingly, Table 2 also tells us that $\Sigma$ is indeed a low-energy state in PZO, with its decrease in energy being about 235.6 meV/f.u. with respect to the cubic state, to be compared with 278.8 meV/f.u. for the known \textit{Pbam} phase. In other words, having finite $\omega_{\Sigma,z}$ and $u_{\Sigma,x}=u_{\Sigma,y}$ (like in the $\Sigma$ and known \textit{Pbam} phases) brings about 85\% of the energy of the known \textit{Pbam} state with respect to $Pm\bar{3}m$  state, with the other 15\% being due to $\omega_{R,x}=\omega_{R,y}$. Such percentages highlight the importance of the bilinear couplings of Eq. (7), which are responsible for the finite values of $\omega_{\Sigma,z}$ and $u_{\Sigma,x}=u_{\Sigma,y}$ in both the $\Sigma$ and known \textit{Pbam} states, as well as their low energies. Table 3 provides the crystal structure and atomic positions of the known \textit{Pbam} and the presently discovered $\Sigma$ phase, respectively, as predicted by direct first-principles calculations at 0\,K, in the hope they can be used in the future to investigate and/or revisit structural phases of PZO.  One can see from Table 3 that the pseudo lattice parameters of the $\Sigma$ phase are not close to each other at 0K, while they are at high temperatures (see Fig 4a).

\section{Conclusions}

We developed and used an original atomistic effective Hamiltonian, incorporating the bilinear coupling discovered in Ref.[\onlinecite{Kinnaryuw}], in order to investigate 
structural and dielectric properties of bulk PZO  as a function of temperature. We also conducted X-ray diffraction measurements for various  temperatures. This effective Hamiltonian reproduces well (i)  the existence of the known low-temperature  AFE orthorhombic \textit{Pbam} phase and the high-temperature paraelectric cubic phase; (2)  the large dielectric response; and (3) the thermal expansion of the pseudocubic lattice parameters of PZO.
Its most provocative prediction is the existence of an intermediate phase  named here as the $\Sigma$ phase and that occurs in-between the known \textit{Pbam} and cubic phases. 
This $\Sigma$ phase  has the \textit{Pbam} symmetry too, therefore yielding an  isostructural  transition in PZO bulks. It mostly differs from the \textit{Pbam} ground state of PZO by the vanishing of anti-phase octahedral tiltings, and its low energy is confirmed by conducting additional direct first-principles calculations. Both the known \textit{Pbam} state and this $\Sigma$ phase are numerically found  to emerge due to the bilinear coupling terms defined in Eqn. (7). Possible reasons explaining why this $\Sigma$ phase may have been previously overlooked are provided here too. They include the possibilities that the  $\Sigma$ phase  only exists in a short-range fashion or is broken into small domains in real materials. Another plausible reason is that its three pseudocubic lattice parameters are basically equal to each other, therefore giving the wrong impression that it is cubic in nature when conducting, {\it e.g.}, diffractions studies. Moreover, it is also possible that  the ferroelectric tetragonal-like phase (that is found to be very close in energy to the $\Sigma$ phase) is experimentally observed, as it might be triggered under external stimulus (electric field, doping/defects or stress) which can thus explain some reported polar intermediate phase or polar clusters far above the known Pbam state. We hope that the present work, predicting a phase that has never been mentioned in the rich literature of the textbook antiferroelectric compound to the best of our knowledge, will generate new experimental, theoretical and computational studies and/or analysis on PZO.

The authors in Arkansas thank the Vannevar Bush Faculty Fellowship (VBFF) Grant No. N00014-20-1-2834 from the Department of Defense and the Office of Naval Research 
Grant No. N00014-21-1-2086. B.X. thanks the National Natural Science Foundation of China under Grant No. 12074277, Natural Science Foundation of Jiangsu Province (BK20201404), the startup fund from Soochow University and the support from Priority Academic Program Development (PAPD) of Jiangsu Higher Education Institutions.
We are also thankful for the computational support from the Arkansas High Performance
Computing Center (AHPCC).
PEJ and RF thank Dr. F. Porcher from Laboratoire L\'eon Brillouin for performing the neutron diffraction and Rietveld analyses.

\newpage

\begin{table*} 
	\centering
		\caption{Parameters of the presently developed effective Hamiltonian for PbZrO$_3$. Atomic units are used and the reference cubic lattice parameter is 7.763 Bohr.}
	\begin{tabular}{l|lr|lr|lr}
		\hline\hline
		Dipole & $Z^*$ & +6.383 & $\epsilon_\infty$ & +6.970 & & \\
		\hline
		$u$ on-site & $\kappa_2$ & +0.00628158 & $\alpha$ & +0.00943 & $\gamma$ & -0.00201 \\
		\hline
		\multirow{3}{*}{$u$ short range} & $j_1$ & -0.004023 & $j_2$ & +0.008550 &  &  \\
		& $j_3$ & +0. 000614& $j_4$ & -0.0005768 & $j_5$ & +0.0004896 \\
		& $j_6$ & +0.0000301 & $j_7$ & +0.0000151  & &  \\
		\hline 
		Elastic & $B_{11}$ &  +4.775 & $B_{12}$ & +1.302 & $B_{44}$ & +0.0.912 \\
		\hline
		$u$-strain coup. & $B_{1xx}$ & -0.293 & $B_{1yy}$ & +0.0522 & $B_{4yz}$ & +0.00294 \\
		\hline
		$\omega$ on-site & $\kappa_A$ & -0.15579& $\alpha_A$ & +3.523804 & $\gamma_A$ & -3.20618 \\
		\hline
		$\omega$ short-range & $k_1$ & +0.0389464& $k_2$ & +0.00413634 & $K'$ & -0.1241 \\
		\hline
		$\omega$-strain coup. & $C_{1xx}$ & -0.317612 & $C_{1yy}$ &+1.440823 & $C_{4yz}$ & -0.011972 \\
		\hline
		$\omega u$ coup. (bilinear)  & $F_{ii}$ & +0.02117 &  & & &  \\
		\hline
		$\omega u$ coup. (trilinear)  & $D_{ii,xy}$ & -0.0430 &  & & &  \\
		\hline
		$\omega u$ coup. (bi-quadratic)  & $E_{xxxx}$ & +0.13945 & $E_{xxyy}$ & +0.339915 & $E_{xyxy}$ & -0.667303 \\
		\hline\hline

	\end{tabular}
	\label{tab1}
\end{table*}

\begin{table}[h]
\caption{Energetic gain for different low-energy states with respect to the cubic paraelectric  $Pm\bar{3}m$ state, as predicted by direct first-principles calculations of PbZrO$_{3}$ }
\begin{tabular}{|c|c|lll}
\cline{1-2}
Phases  &    Energy(meV/fu)  &  &  &  \\
\cline{1-2}
\textit{Pbam}      &278.8     &  &  &  \\ \cline{1-2}
$R3c$   & 277.7    &  &  &  \\ \cline{1-2}
$\Sigma$  & 235.6    &  &  &  \\ \cline{1-2}
\end{tabular}
\end{table}

{
\begin{table}[h]
\caption{Crystal structure and  Atomic positions of the Pbam and $\Sigma$ phases at 0K obtained from first-principles calculations}
\begin{tabular}{|cccccclc|}
\hline
\multicolumn{8}{|c|}{Pbam    phase}                                                                                                                                                                                                                                                                                                                   \\ \hline
\multicolumn{3}{|c|}{Lattice Constants}                                                                                                          & \multicolumn{5}{c|}{Cell angles}                                                                                                                                                                   \\ \hline
\multicolumn{1}{|c|}{a}                      & \multicolumn{1}{c|}{b}                       & \multicolumn{1}{c|}{c}                             & \multicolumn{1}{c|}{$\alpha$}                       & \multicolumn{1}{c|}{$\beta$} & \multicolumn{3}{c|}{$\gamma$}                                                                                 \\ \hline
\multicolumn{1}{|c|}{11.782}                 & \multicolumn{1}{c|}{5.887}                   & \multicolumn{1}{c|}{8.188}                         & \multicolumn{1}{c|}{90}                             & \multicolumn{1}{c|}{90}      & \multicolumn{3}{c|}{90}                                                                                       \\ \hline
\multicolumn{1}{|c|}{\multirow{2}{*}{Label}} & \multicolumn{1}{c|}{\multirow{2}{*}{Symbol}} & \multicolumn{1}{c|}{\multirow{2}{*}{Multiplicity}} & \multicolumn{1}{c|}{\multirow{2}{*}{Wyckoff label}} & \multicolumn{3}{c|}{Fractional coordinates}                                                & \multicolumn{1}{l|}{\multirow{2}{*}{Occupancy}} \\ \cline{5-7}
\multicolumn{1}{|c|}{}                       & \multicolumn{1}{c|}{}                        & \multicolumn{1}{c|}{}                              & \multicolumn{1}{c|}{}                               & \multicolumn{1}{c|}{x}       & \multicolumn{1}{c|}{y}       & \multicolumn{1}{c|}{z}       & \multicolumn{1}{l|}{}                           \\ \hline
\multicolumn{1}{|c|}{Pb1}                    & \multicolumn{1}{c|}{Pb}                      & \multicolumn{1}{c|}{4}                             & \multicolumn{1}{c|}{g}                              & \multicolumn{1}{c|}{0.87505} & \multicolumn{1}{l|}{0.20121} & \multicolumn{1}{l|}{0.00000} & 1                                               \\ \hline
\multicolumn{1}{|c|}{Pb2}                    & \multicolumn{1}{c|}{Pb}                      & \multicolumn{1}{c|}{4}                             & \multicolumn{1}{c|}{h}                              & \multicolumn{1}{c|}{0.87134} & \multicolumn{1}{l|}{0.20890} & \multicolumn{1}{l|}{0.50000} & 1                                               \\ \hline
\multicolumn{1}{|c|}{Zr1}                    & \multicolumn{1}{c|}{Zr}                      & \multicolumn{1}{c|}{8}                             & \multicolumn{1}{c|}{i}                              & \multicolumn{1}{c|}{0.62378} & \multicolumn{1}{l|}{0.24246} & \multicolumn{1}{l|}{0.24995} & 1                                               \\ \hline
\multicolumn{1}{|c|}{O1}                     & \multicolumn{1}{c|}{O}                       & \multicolumn{1}{c|}{4}                             & \multicolumn{1}{c|}{e}                              & \multicolumn{1}{c|}{0.00000} & \multicolumn{1}{l|}{0.00000} & \multicolumn{1}{l|}{0.20287} & 1                                               \\ \hline
\multicolumn{1}{|c|}{O2}                     & \multicolumn{1}{c|}{O}                       & \multicolumn{1}{c|}{4}                             & \multicolumn{1}{c|}{f}                              & \multicolumn{1}{c|}{0.00000} & \multicolumn{1}{l|}{0.50000} & \multicolumn{1}{l|}{0.77054} & 1                                               \\ \hline
\multicolumn{1}{|c|}{O3}                     & \multicolumn{1}{c|}{O}                       & \multicolumn{1}{c|}{8}                             & \multicolumn{1}{c|}{i}                              & \multicolumn{1}{c|}{0.76195} & \multicolumn{1}{l|}{0.03295} & \multicolumn{1}{l|}{0.28113} & 1                                               \\ \hline
\multicolumn{1}{|c|}{O4}                     & \multicolumn{1}{c|}{O}                       & \multicolumn{1}{c|}{4}                             & \multicolumn{1}{c|}{h}                              & \multicolumn{1}{c|}{0.09438} & \multicolumn{1}{l|}{0.19920} & \multicolumn{1}{l|}{0.50000} & 1                                               \\ \hline
\multicolumn{1}{|c|}{O5}                     & \multicolumn{1}{c|}{O}                       & \multicolumn{1}{c|}{4}                             & \multicolumn{1}{c|}{g}                              & \multicolumn{1}{c|}{0.15771} & \multicolumn{1}{l|}{0.22353} & \multicolumn{1}{l|}{0.00000} & 1                                               \\ \hline
\multicolumn{8}{|c|}{$\Sigma$    phase}                                                                                                                                                                                                                                                                                                               \\ \hline
\multicolumn{3}{|c|}{Lattice Constants}                                                                                                          & \multicolumn{5}{c|}{Cell angles}                                                                                                                                                                   \\ \hline
\multicolumn{1}{|c|}{a}                      & \multicolumn{1}{c|}{b}                       & \multicolumn{1}{c|}{c}                             & \multicolumn{1}{c|}{$\alpha$}                       & \multicolumn{1}{c|}{$\beta$} & \multicolumn{3}{c|}{$\gamma$}                                                                                 \\ \hline
\multicolumn{1}{|c|}{11.717}                 & \multicolumn{1}{c|}{5.913}                   & \multicolumn{1}{c|}{4.076}                         & \multicolumn{1}{c|}{90}                             & \multicolumn{1}{c|}{90}      & \multicolumn{3}{c|}{90}                                                                                       \\ \hline
\multicolumn{1}{|l|}{\multirow{2}{*}{Label}} & \multicolumn{1}{l|}{\multirow{2}{*}{Symbol}} & \multicolumn{1}{l|}{\multirow{2}{*}{Multiplicity}} & \multicolumn{1}{l|}{\multirow{2}{*}{Wyckoff label}} & \multicolumn{3}{c|}{Fractional coordinates}                                                & \multicolumn{1}{l|}{Occupancy}                  \\ \cline{5-8} 
\multicolumn{1}{|l|}{}                       & \multicolumn{1}{l|}{}                        & \multicolumn{1}{l|}{}                              & \multicolumn{1}{l|}{}                               & \multicolumn{1}{c|}{x}       & \multicolumn{1}{c|}{y}       & \multicolumn{1}{c|}{z}       & \multicolumn{1}{l|}{}                           \\ \hline
\multicolumn{1}{|c|}{Pb1}                    & \multicolumn{1}{c|}{Pb}                      & \multicolumn{1}{c|}{4}                             & \multicolumn{1}{c|}{g}                              & \multicolumn{1}{c|}{0.64479} & \multicolumn{1}{c|}{0.30421} & \multicolumn{1}{c|}{0.00000} & 1                                               \\ \hline
\multicolumn{1}{|c|}{Zr1}                    & \multicolumn{1}{c|}{Zr}                      & \multicolumn{1}{c|}{4}                             & \multicolumn{1}{c|}{h}                              & \multicolumn{1}{c|}{0.61911} & \multicolumn{1}{c|}{0.75709} & \multicolumn{1}{c|}{0.50000} & 1                                               \\ \hline
\multicolumn{1}{|c|}{O1}                     & \multicolumn{1}{c|}{O}                       & \multicolumn{1}{c|}{4}                             & \multicolumn{1}{c|}{h}                              & \multicolumn{1}{c|}{0.73631} & \multicolumn{1}{c|}{0.45735} & \multicolumn{1}{c|}{0.50000} & 1                                               \\ \hline
\multicolumn{1}{|c|}{O2}                     & \multicolumn{1}{c|}{O}                       & \multicolumn{1}{c|}{2}                             & \multicolumn{1}{c|}{d}                              & \multicolumn{1}{c|}{0.00000} & \multicolumn{1}{c|}{0.50000} & \multicolumn{1}{c|}{0.50000} & 1                                               \\ \hline
\multicolumn{1}{|c|}{O3}                     & \multicolumn{1}{c|}{O}                       & \multicolumn{1}{c|}{2}                             & \multicolumn{1}{c|}{b}                              & \multicolumn{1}{c|}{0.00000} & \multicolumn{1}{c|}{0.00000} & \multicolumn{1}{c|}{0.50000} & 1                                               \\ \hline
\multicolumn{1}{|c|}{O4}                     & \multicolumn{1}{c|}{O}                       & \multicolumn{1}{c|}{4}                             & \multicolumn{1}{c|}{g}                              & \multicolumn{1}{c|}{0.35576} & \multicolumn{1}{c|}{0.28658} & \multicolumn{1}{c|}{0.00000} & 1                                               \\ \hline
\end{tabular}
\end{table}
}
\end{document}